# Tensile Strains Give Rise to Strong Size Effects for Thermal Conductivities of Silicene, Germanene and Stanene

Youdi Kuang,*[†] Lucas Lindsay,[‡] Sanqiang Shi,[†] Guangping Zheng[†]

[†]Department of Mechanical Engineering, the Hong Kong Polytechnic University, Hung Hom, Kowloon, Hong Kong

[‡]Materials Science and Technology Division, Oak Ridge National Laboratory, Oak Ridge, Tennessee 37831, USA

ABSTRACT: Based on first principles calculations and self-consistent solution of linearized Boltzmann-Peierls equation for phonon transport approach within a three-phonon scattering framework, we characterize lattice thermal conductivities $k$ of freestanding silicene, germanene and stanene under different isotropic tensile strains and temperatures. We find a strong size dependence of $k$ for silicene with tensile strain, i.e., divergent $k$ with increasing system size, in contrast, the intrinsic room temperature $k$ for unstrained silicene converges with system size to 19.34 W/m·K by 178 nm. The room temperature $k$ of strained silicene becomes as large as that of bulk silicon by 84 μm, indicating the possibility of using strain in silicene to manipulate $k$ for thermal management. The relative contribution to the intrinsic $k$ from out-of-plane acoustic modes is largest for unstrained silicene, ~39% at room temperature. The single mode relaxation time approximation, which works reasonably well for bulk silicon, fails to appropriately describe phonon thermal transport in silicene, germanene and stanene within the temperature range considered. For large samples of silicene, $k$ increases with tensile strain, peaks at ~7% strain and then decreases with further strain. In germanene and stanene increasing strain hardens and stabilizes long wavelength out-of-plane acoustic phonons, and leads to similar $k$ behaviors to those of silicene. These findings further our understanding of phonon dynamics in group-IV buckled monolayers and may guide transfer and fabrication techniques of these freestanding samples and engineering $k$ by size and strain for applications of thermal management and thermoelectricity.

"Cousins" of graphene, group-IV monolayer silicene, germanene and stanene were recently fabricated on supporting substrates in 2012,[1] 2014[2] and 2015,[3] respectively. Unlike flat graphene, silicene, germanene and stanene are buckled sheets composed of Si, Ge and Sn atoms, respectively, but maintain the same hexagonal symmetry as graphene. It is widely believed that they are of great promise for applications of thermoelectricity and nanoelectronics.[3] Previous work[4-6] demonstrated that tensile strain effects on electronic properties of these buckled monolayer systems may greatly benefit these applications, however, the effects of strain on thermal transport remain unclear. This is surprising given that the conduction of heat plays a critical role in these applications. Thus, understanding thermal transport in these buckled monolayers is critically important, and of fundamental interest.

In contrast to the extensive experimental and theoretical studies on thermal transport of graphene,[7-14] experimental measurements of thermal conductivities $k$ and systematic theoretical investigations of thermal transport behaviors for these buckled monolayers are still lacking. To date, there is limited theoretical work for silicene, with significant discrepancies among the corresponding results: previous molecular dynamics simulations[15, 16] gave contradictory results of out-of-plane phonon mode contributions, indicating the calculated $k$ strongly depends on the quality of the empirical potentials employed, as argued recently by Xie et al.[17] and Gu and Yang.[18] Using first principles lattice dynamics calculations, Xie et al.[17] and Gu and Yang[18] gave contradictory conclusions on the convergence of intrinsic $k$ of unstrained silicene at room temperature with increasing system size based on the single mode relaxation time approximation (SMRTA) solution[8] of the Boltzmann-Peierls equation (BPE) for phonons. We note that the later also considered a fully iterative solution[13] of the BPE and argued the intrinsic $k$ of unstrained silicene diverges. Both of these calculations demonstrated that the out-of-plane acoustic phonon branch was linear at low frequencies, very different from the familiar quadratic nature for the corresponding branch in unstrained graphene. This may significantly affect the $k$ convergence behavior, as discussed below. Despite the excellent insights into the thermal transport behavior of silicene developed in these previous work, it is clear from the literature that fundamental problems on thermal transport in these monolayer systems remain unclear, including: 1) the role of out-of-plane acoustic phonon modes; 2) the validity of the SMRTA; 3) the effects of size, strain and temperature and their interplay.

In this work, we use a rigorous first principles BPE for phonon transport approach which has accurately predicted $k$ of graphene at different sizes and temperatures,[19] to systemically investigate thermal transport behaviors in unstrained and strained group-IV buckled monolayers. We find that the SMRTA does not appropriately describe phonon thermal transport in these systems, and tensile strain, which stabilizes the freestanding germanene and stanene lattices, gives strong size effects of $k$. Further, out-of-plane acoustic phonon modes play an important role for thermal transport, though not nearly as important as in graphene.

The computational methodology is as follows: a microscopic description of $k$ can be derived from the BPE for phonons within the three-phonon scattering framework. The intrinsic $k$ for an infinite monolayer is isotropic and is given by $k = k^{\alpha\alpha} = \dfrac{1}{k_{\mathrm{B}}T^2 V N_1^2} \displaystyle\sum_{\lambda} f_\lambda (f_\lambda + 1)(\hbar\omega_\lambda)^2 v_\lambda^\alpha v_\lambda^\alpha \tau_\lambda$,[13, 20] where $\omega_\lambda$, $v_\lambda^\alpha$, and $\tau_\lambda$ are the angular frequency, group velocity and phonon lifetime, respectively, and $\alpha$ is an in-plane crystallographic direction. Here $\lambda$ represents a phonon mode denoted by its wavevector and branch index. $k_{\mathrm{B}}$, $\hbar$, $f_\lambda$ are the Boltzmann constant, the reduced Plank constant and the Bose-Einstein distribution of phonons at temperature $T$, respectively. $V$ is the volume of the monolayer unit cell with a thickness $h$. To be consistent with a previous definition for the thickness of graphene[13] and silicene[15-18] for $k$ calculations, these monolayer thicknesses are taken as the equilibrium van der Waals interaction distance: $h$-silicene=4.20 Å, $h$-germanene=4.22 Å and $h$-stanene=4.34 Å. The full 2D first Brillouin zone is discretized into a $\Gamma$-centered regular $N_1 \times N_1$ grid with $N_1$ up to 301 with the smallest wavevector $q=2\pi/(a \times N_1)$ used in this study, here $a$ is the in-plane lattice constant for a given strain. This work combines an iteratively self-consistent solution to the linearized BPE with harmonic and anharmonic interatomic force constants (IFCs) from density functional theory (DFT) calculations using the VASP package[21] within the local density approximation and potentials given by the projector augmented wave method. To determine the harmonic IFCs, DFT calculations are employed with a $9 \times 9$ supercell without additional neighbor cutoff, a $3 \times 3$ k-point mesh and a single-point energy convergence precision $1 \times 10^{-10}$ eV for all of the 2-atom unit cells considered. The plane-wave cutoff energy is taken as 480 eV, 340 eV and 180 eV respectively for silicene, germanene and stanene. The corresponding vacuum layer thickness is taken as 10 Å, 16 Å and 20 Å, respectively. To determine the anharmonic IFCs, DFT calculations with $\Gamma$-point sampling in slightly perturbed $5 \times 5$ supercells with a $5 \times 5$ k-point mesh and the same plane-wave cutoff to those for the corresponding harmonic IFCs are conducted. Interactions are considered out to sixth nearest neighbors (NN), corresponding to about 1 nm, of the unit cell atoms. Tests based on an $8 \times 8$ supercell of unstrained silicene and interaction cutoffs from second NN up to tenth NN demonstrate that sixth NN interactions are required to give converged anharmonic IFCs. We used sixth NN anharmonic interactions for all of the systems studied here. Translational invariance conditions are enforced on all of the IFCs. Further technical details for the calculations of dispersion relations by the finite displacement approach and calculation of the three-phonon scattering rates can be found elsewhere.[20, 22] Before calculating IFCs, all unit cells considered are relaxed fully using a 27 × 27 k-point mesh, an energy convergence precision $1 \times 10^{-10}$ eV, a force convergence precision $1 \times 10^{-6}$ eV/Å and the same plane-wave cutoff and vacuum layer thickness as those for the IFC calculations. The obtained equilibrium lattice constant $a_0$ and buckling height $h_0$, [23, 24] vertical distance between the two neighbor atoms of the 2-atom unit cells for each system are: $a_0$-silicene=3.825 Å, $a_0$-germanene=3.968 Å, $a_0$-stanene=4.558 Å; $h_0$-silicene=0.439 Å, $h_0$-germanene=0.647 Å, $h_0$-stanene=0.808 Å. These values agree well with previous first principles calculations.[23, 24]

We first calculate the phonon dispersion relations for silicene, germanene and stanene under different isotropic tensile strains denoted by $\varepsilon$ ($\varepsilon \leq 0.1$), $\varepsilon = (a-a_0)/a_0$. Increasing strain decreases the buckling height due to the Poisson effect[25] in each monolayer. Figure 1 shows the dispersion results of three representative strain levels for each monolayer system. Here, for consistency with the description of graphene, out-of-plane acoustic modes are labeled " ZA " modes though they have some coupling with the in-plane longitudinal and transversal acoustic (LA and TA) modes due to the buckled atomic configuration. For unstrained silicene, i.e., $\varepsilon = 0$, it can be seen from Figure 1a that the ZA dispersion is quadratic-like but not purely quadratic due to the loss of reflection symmetry. Numerical fitting gives the approximate relation $\omega \sim q^x$ ($x=1.6$) between frequency $\omega$ and wavevector magnitude $q$ for the ZA branch considering the entire frequency range. Physically, as the in-plane size goes to infinity, the thickness effect (buckling) on the dispersion which results in coupling of in-plane and out-of-plane displacements of a thin plate disappears because the ratio of buckling height to in-plane size goes to zero. Therefore, based on previous elastic theory,[25] for an infinite monolayer, the group velocity is zero at the $\Gamma$-point and the dispersion relation is quadratic ($x$=2). When the in-plane size is comparable to the buckling height, the corresponding system looks bulk-like and the corresponding ZA dispersion is expected to be nearly linear due to this coupling. Thus, with increasing in-plane size $x$ will gradually increase to 2 at the $\Gamma$-point. Correspondingly, the group velocity of the ZA mode gradually goes to zero with decreasing $q$ magnitude. We note that from elasticity theory it is not possible to have a ZA branch with zone center linear dispersion and the constant group velocity in the long wavelength limit in unstrained monolayer systems. A constant group velocity leads to a constant out-of-plane Young's modulus[26] and sound velocity, which violates the basic fact that relaxed atomic monolayers do not have these. With increasing strain the ZA modes harden, increasing linearization of the ZA branch similar to that of strained graphene,[8, 13] giving a constant group velocity component in the long wave limit once tensile strain is applied. This is consistent with the rigorous analytical predictions from continuum mechanics theory for a thin plate.[25]

Phonon softening is seen for all longitudinal, transversal and flexural optical modes (LO, TO and ZO, respectively) in the strained systems. These variations are similar to those in flat graphene and monolayer hexagonal boron nitride.[19, 22] However, unlike the softening of LA and TA modes found in graphene, the corresponding dispersion changes only slightly with increasing strain in silicene, especially at low strain. Unlike flat graphene with no mixing of in-plane and out-of-plane modes, these modes are coupled in buckled systems, not purely in- or out-of-plane. Thus, as in-plane components of the LA and TA soften they are compensated by out-of-plane components causing only slight changes.

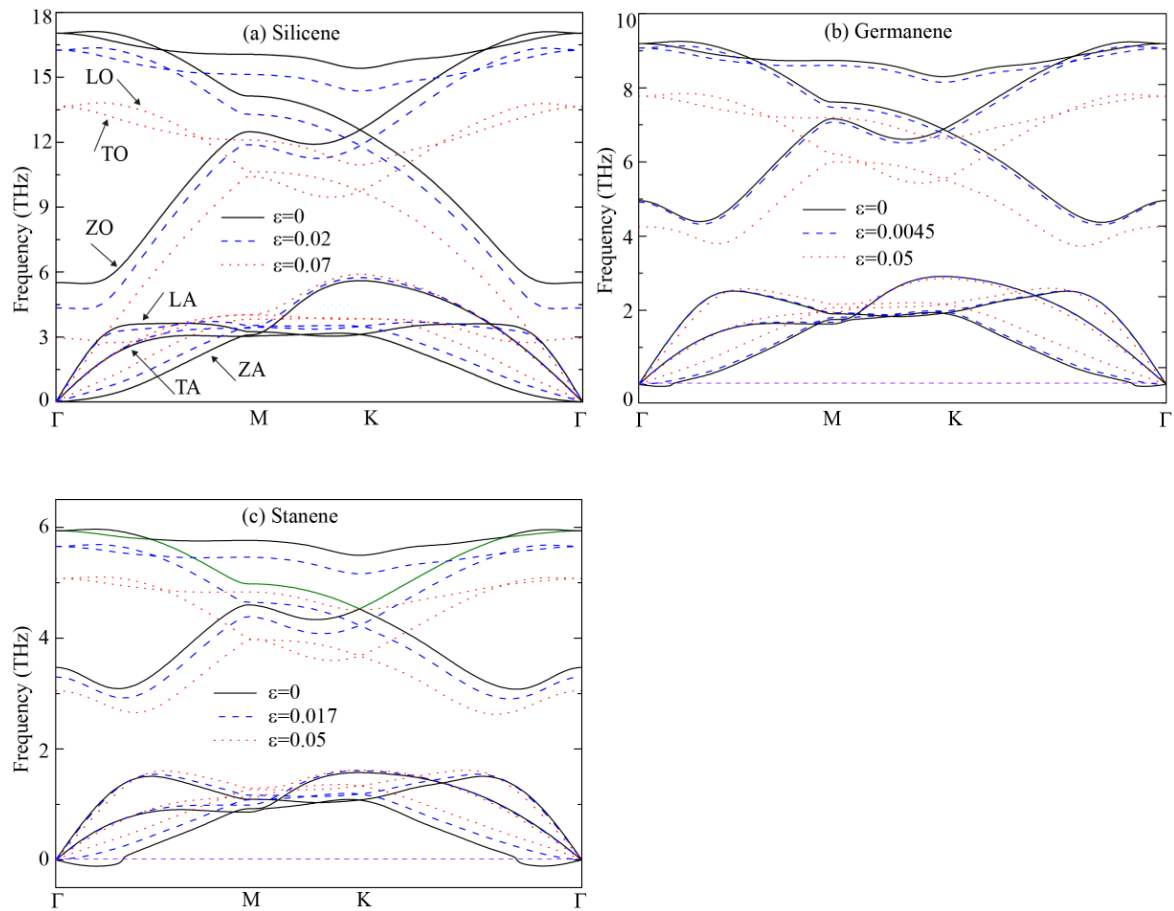

Figure 1. Calculated phonon dispersions of silicene (a), germanene (b) and stanene (c) for different tensile strains.

Unlike silicene, calculations of the ZA dispersion of unstrained germanene and stanene give imaginary frequencies, i.e, soft ZA modes in the vicinity of the zone center, indicating that these freestanding periodic lattices are not stable. Previous first principles calculations[23, 27] of the dispersion based on both interpolation methods and linear response theory for germanene gave similar results using different potentials and computational packages, demonstrating that the imaginary frequencies are likely not a result of numerical error. We also considered *d*-electrons in the valence shell for our dispersion calculations (see Supplementary Figure S1). These calculations also give imaginary modes for unstrained germanene, further verifying the predicted instability of this freestanding monolayer structure. Previous nonlinear vibration theory of a thin plate with an initial out-of-plane deformation analytically showed[28, 29] that the plate will lose vibrational stability in the form of dynamical buckling, especially at low frequencies, consistent with the soft ZA modes observed here with initial buckling height above a critical value. Such critical values depend on the mechanical boundary conditions[29] (typically supported or clamped boundary conditions) of suspended finite samples. These results may partly explain why large germanene and stanene samples were fabricated previously only on substrates.[2, 3] The reason is that the interfacial van der Waals interaction may harden the ZA modes[30] and help stabilize them, similar to what the tensile stain does here. We conduct a series of dispersion calculations on strained germanene and stanene with small strain increments to search for their stable states. As shown in Figure 1b and 1c, the soft modes are removed at critical strain values 0.0045 and 0.017, respectively, due to ZA mode hardening. The corresponding ZA dispersion relations are not purely quadratic, similar to that of unstrained silicene. Variations in the phonon dispersions of other modes in germanene and stanene are similar to those of silicene with increasing strain. The present results suggest that large freestanding germanene and stanene samples may be achieved if appropriate tensile strains are maintained.

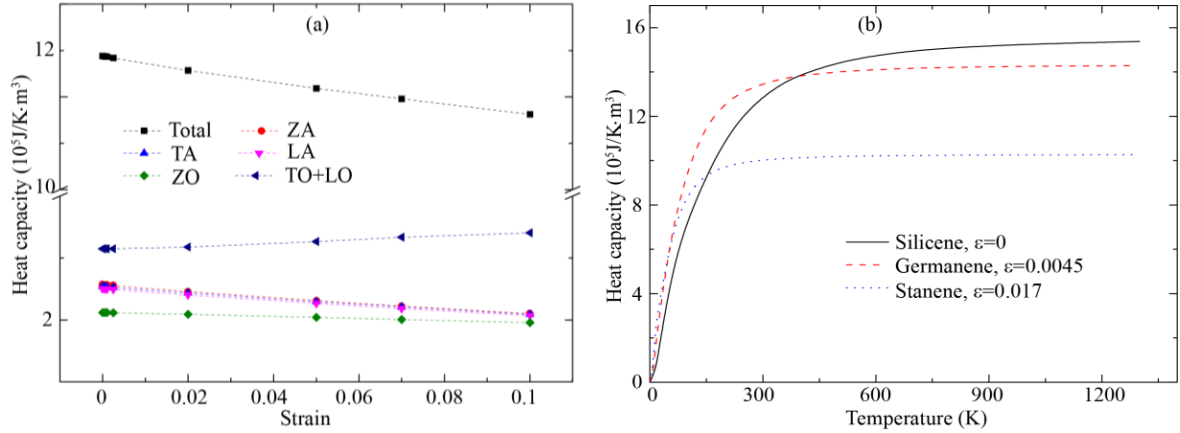

Figure 2. (a) Strain dependent heat capacity (and mode contributions) of silicene at room temperature. (b) Temperature dependent heat capacities of unstrained silicene, germanene at $\varepsilon$=0.0045 and stanene at $\varepsilon$=0.017 (critical strain values).

The heat capacity is an important component of $k$. Figure 2a shows the strain dependent room temperature heat capacity and corresponding branch contributions for silicene. It can be seen that the total heat capacity decreases with increasing strain due to decreasing contributions from most modes including ZA, TA, LA and ZO. This result is different from that of graphene[19] for which the total heat capacity decreases first at small strains and then increases mainly due to the competition between the decreased ZA mode contribution and the increased contribution from optical modes with increasing stain. The decrease of the ZA mode contributions for silicene and graphene is due to the decreased phonon density of states from phonon hardening and linearization. Figure 2b shows the temperature dependent heat capacities for unstrained silicene, germanene at $\varepsilon$ =0.0045 and stanene at $\varepsilon$ =0.017 (critical strains). Similar temperature dependence is seen. This data may be used to estimate the heat capacities of the corresponding multilayer systems[3, 31, 32] and layered bulk systems at different temperatures.

In the following, we will discuss strain, temperature and size dependent lattice $k$ of the stable monolayer systems.

For each system (strained and unstrained), the convergence[22] of $k$ is tested as a function of $N_1$ with a precision of $1 \times 10^{-5}$ difference between iterative steps. This ensures full self-consistent convergence of $k$ (see Supplementary Figure S2). Room temperature $k$ convergence tests corresponding to three representative strain levels for each monolayer system are shown in Figure 3. For unstrained silicene (Figure 3a), germanene (Figure 3b) and stanene (Figure 3c) at the corresponding critical strains, the intrinsic room temperature $k$ converge at values 19.34 W/m·K, 10.52 W/m·K and 1.25 W/m·K, respectively. Natural isotopes[13, 33] decrease the corresponding intrinsic $k$ by 2.1%, 5.5% and 6.9%, respectively. It is not surprising that the intrinsic $k$ of these monolayer systems converges considering the most recently reported $k$ [10, 22] for other buckled monolayers including graphane, fluorographene and molybdenum disulphide[7] and flat monolayers including graphene and boron nitride[19] were also reported to converge. For each *strained* case, $k$ increases approximately logarithmically with increasing $N_1$, indicating that $k$ diverges logarithmically with increasing sample size. The corresponding SMRTA solutions of *k* show the same trends. We note that strained graphene also has divergent *k* ,[8, 22] indicating the strain dependent convergence behaviors in *k* are inherent in the three-phonon scattering framework for group-IV monolayers. Given the similarities in *k* behaviors of these buckled monolayer systems we focus our further discussion below on silicene.

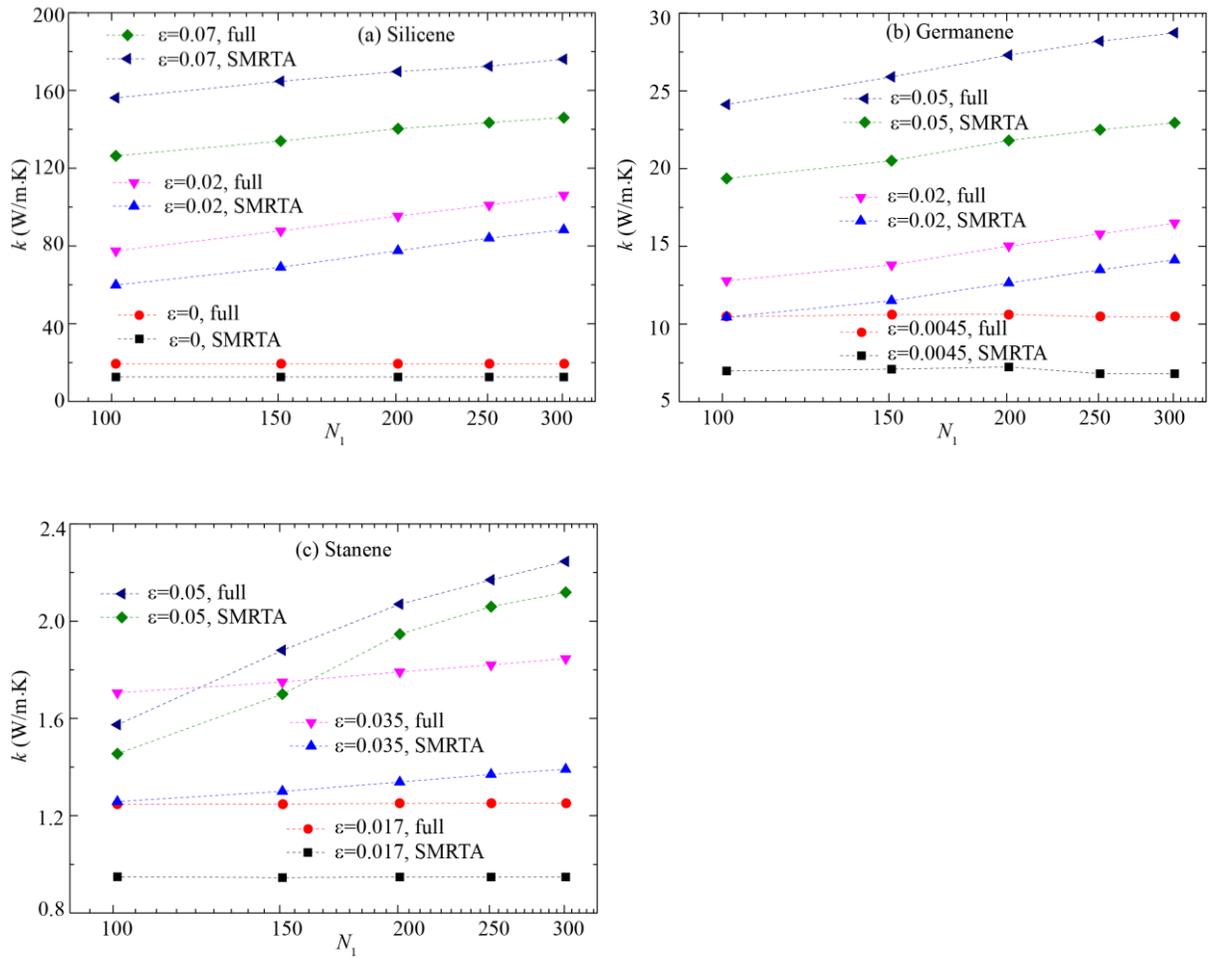

Figure 3. Intrinsic *k* with respect to *q*-point grid density $N_1$ for silicene (a), germanene (b) and stanene (c) for different strains at room temperature.

To further verify the strain dependent convergence trends, especially at small strains, we plot the calculated saturated mean free path (SMFP) for a given $N_1$ with respect to $N_1$ in Figure 4 for silicene at strains $\varepsilon=0$ and $\varepsilon=0.0005$. The SMFP is defined as the largest MFP of all phonon modes considered at a given $N_1$. It can be seen that the SMFP tends to converge for $\varepsilon=0$ (Figure 4a) while diverges $\varepsilon=0.0005$ (Figure 4b), consistent with the convergence trends of $k$ for $\varepsilon=0$ and $\varepsilon=0.0005$, respectively. The SMFP from ZA modes converges to 178 nm, 184 nm and 16 nm, respectively for unstrained silicene, germanene and stanene at the corresponding critical strains, indicating insignificant size effects of $k$ for typical system sizes. Critically strained stanene has one order of magnitude lower SMFP than those of critically strained germanene, which correlates strongly with the significantly lower $k$ of stanene. A corresponding comparison shows more than one order of magnitude larger scattering rates of ZA modes of critically strained stanene than those of critically strained germanene. We note that recent work[34] reported a much higher $k=\sim 11$ W/m·K for unstrained stanene likely because of the corresponding unphysical linear ZA dispersion used by the authors. The behavior of $k$ of unstrained graphene is quite different as a strong size effect of $k$ has been observed experimentally[14] and predicted up to ~8 cm theoretically.[19] The strong size effect on $k$ for the strained buckled group-IV monolayers results from the divergence of $k$ with tensile strains. For LA and TA modes having linear dispersions, the corresponding phonon density of states tend to zero in the long wavelength limit, their contributions do not lead to the $k$ divergence. As seen from the formula for $k$ calculations above, both the nonzero group velocity component of the ZA modes of strained silicene and the divergent SMFP or lifetime (see Figures 4b and 6b) lead to nonzero contributions of ZA modes in the long wavelength limit. Therefore, the long wavelength ZA modes are responsible for the $k$ divergence and the strong size effect. We note that the transition from convergence to divergence of $k$ with application of strain is abrupt due to the abrupt transition from ZA dispersion lacking a linear component in the long wavelength limit to ZA dispersion with a linear component. This abrupt dispersion transition in the long wavelength limit has been shown analytically by the elastic theory of a thin plate.[25, 29] To clearly observe the $k$ divergence behavior in lattice dynamics calculations, a series of large $q$-point grid densities ($N_1 \geq 100$) are required to sample long wavelength ZA modes at low frequencies.

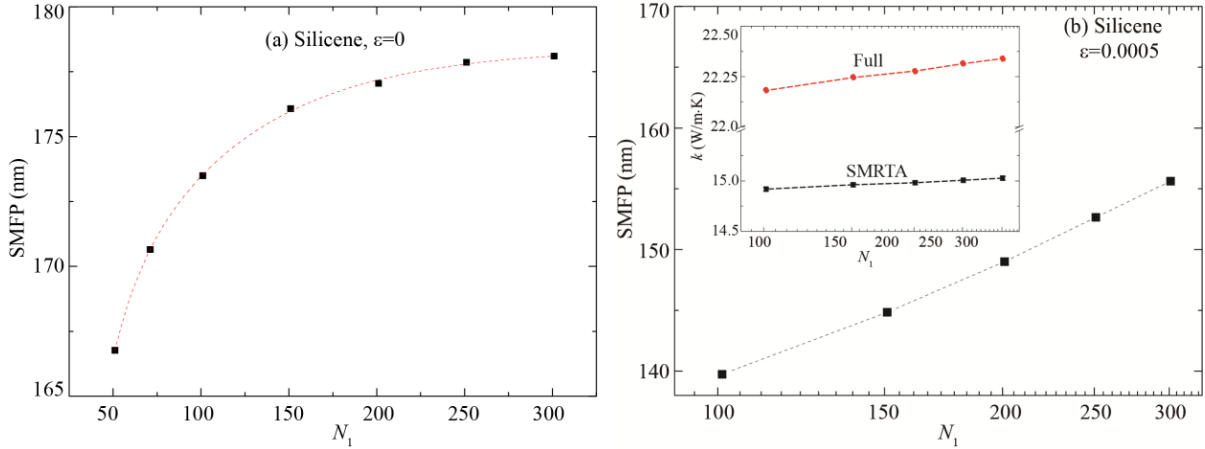

Figure 4. Saturated MFP with respect to $q$-point grid density $N_1$ for silicene at $\varepsilon=0$ (a) and $\varepsilon=0.0005$ (b) at room temperature. The inset shows the corresponding convergence of $k$ with respect to $N_1$.

Figure 5a shows the temperature effect on the convergence of $k$ for $\varepsilon=0.05$. The intrinsic $k$ diverges at lower temperature (80 K) and higher temperature (800 K) from both the full and SMRTA solutions, indicating $k$ divergence of strained systems is inherent in three-phonon scattering theory framework and the strong size effect is fairly temperature independent. The convergence of $k$ for these strained buckled systems with consideration of higher order scattering (e.g., from four phonon processes) is an open question[8] and beyond the scope of this work. Figure 5b shows the temperature dependent ratio of intrinsic $k$ by full solution to that by the SMRTA solution ($k_{full}/k_{SMRTA}$) and the weight of the corresponding branch contributions to the full-solution of $k$ ($k_{mode}/k_{full}$) in unstrained silicene. The $k_{full}/k_{SMRTA}$ ratio is very large at low temperature and decreases to a constant value ~1.48 with increasing temperature. Similar behavior is seen for germanene and stanene at their corresponding critical strains. This indicates that the SMRTA fails to give an appropriate description of phonon thermal transport in group-IV monolayers, i.e, Normal processes (phonon momentum strictly conserved) play a critical role in phonon redistribution even at high temperatures where the resistive Umklapp scattering is expected to be more important. The ZA mode contributions decrease with increasing temperature but still contribute ~34% at high temperatures, twice that of the TA or LA modes, indicating ZA modes play an important role for thermal transport, though significantly less than in graphene. Unlike graphene for which the optical modes give negligible contributions to $k$, the ZO modes give significant contributions to $k$ of silicene for $T > 300K$ mainly due to the large group velocity of the ZO modes (see Figure 1). For strained room temperature silicene, the $k_{full}/k_{SMRTA}$ ratios at different strains show the SMRTA still fails to give an appropriate description of phonon thermal transport, as seen in Figure 5c. We note that this ratio depends slightly on $N_1$ from our calculations, as seen from Figures 4 and 5a. Moreover, as an example, using larger $N_1=401$ gives a ratio variation of only 0.8% as $N_1$ increases from 301 to 401 for $\varepsilon=0.05$. The mode contributions from different acoustic branches are comparable except that the ZA mode contribution is dominant at the largest strain $\varepsilon=0.1$. However, for strained graphene[8, 19] the ZA mode contributions dominate for all $\varepsilon \leq 0.1$.

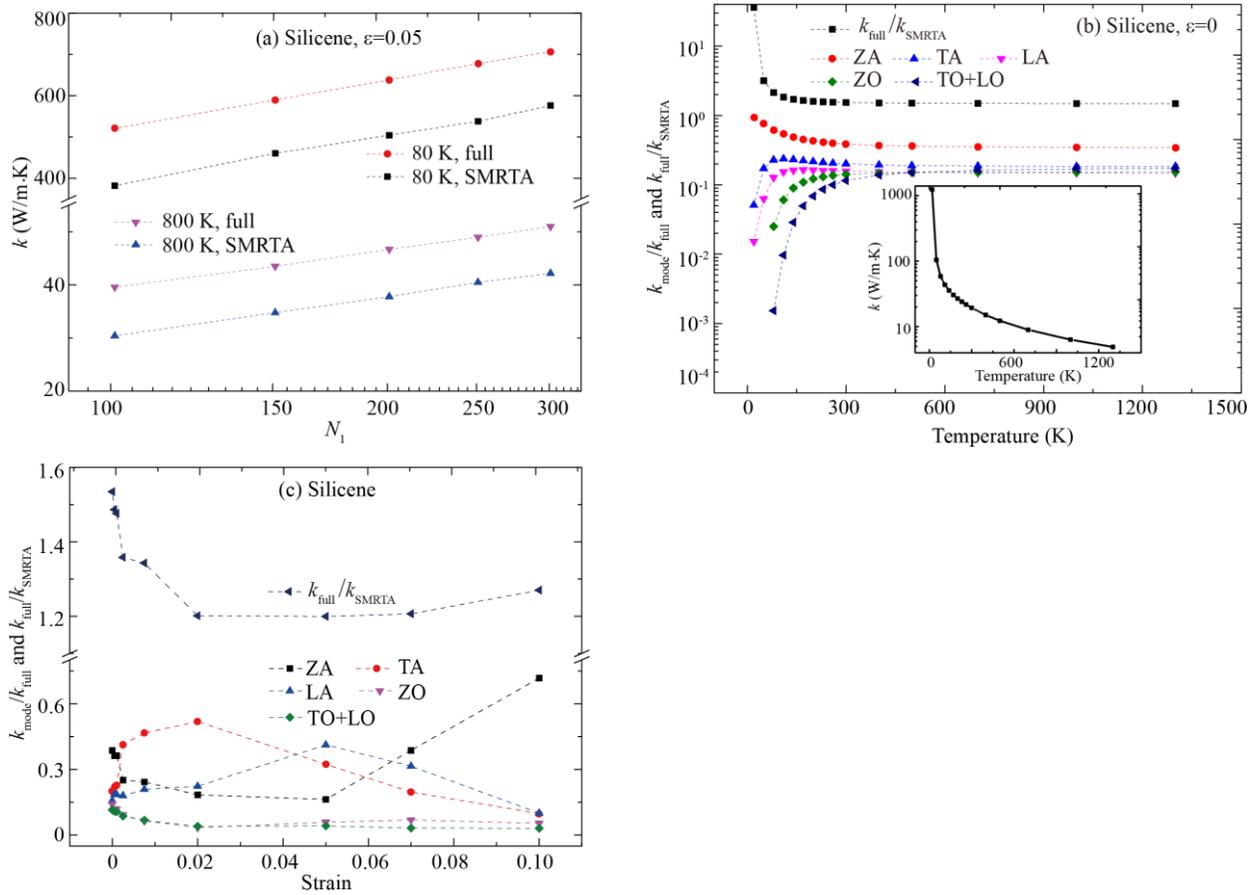

Figure 5. (a) Intrinsic $k$ with respect to $q$-point grid density $N_1$ for strained silicene at different temperatures. (b) $k_{full}/k_{SMRTA}$ and mode contributions ($k_{mode}/k_{full}$) versus temperature for unstrained silicene. The inset shows the temperature dependent intrinsic $k$ of unstrained silicene. (c) $k_{full}/k_{SMRTA}$ and $k_{mode}/k_{full}$ versus strain for silicene at room temperature with $N_1$=301.

For *finite* unstrained and strained monolayers, the corresponding scattering rate $\tau_\lambda^f$ is calculated using the Matthiessen rule[20] expressed here as $\frac{1}{\tau_\lambda^f} = \frac{1}{\tau_\lambda} + \frac{1}{\tau_\lambda^b}$. $\frac{1}{\tau_\lambda}$ is the intrinsic phonon-phonon scattering rate; $\frac{1}{\tau_\lambda^b}$ represents the scatting rate by contact boundaries and is expressed empirically as $\frac{1}{\tau_\lambda^b} = \frac{2|v_\lambda^x|}{L}$,[13] where $v_\lambda^x$ is the group velocity along the thermal transport direction and $L$ is the sample length. Figure 6a shows the intrinsic room temperature $k$ corresponding to $N_1$=301 for infinite samples of silicene, germanene and stanene at different strains. The peak enhancement in $k$ can be seen at strain ~0.07, ~0.05 and ~0.05, respectively. Such behavior is also observed for smaller $N_1$, implying there is a peak enhancement in $k$ with increasing strain for finite monolayer systems. We note that such peak enhancement is also seen in multilayer graphene and graphite.[22] Based on the $\frac{1}{\tau_\lambda}$ corresponding to $N_1$=301 at the peak strain and tests on a series of $L$ values, we obtain nearly the same peak $k$ to those in Figure 6a for finite samples with $L$ values 84 μm, 56 μm and 23.6 μm for silicene, germanene and stanene, respectively. The corresponding boundary scattering rates are far smaller than the intrinsic phonon-phonon scattering rates at these sizes. The peak enhancement mainly results from the competition between the strain dependent heat capacity and lifetimes of acoustic modes. The strain dependent lifetimes of ZA, TA and LA modes are shown in Figure 6b, 6c and 6d, respectively, for silicene with various strains. For $\varepsilon \leq 0.07$ the lifetime increases with increasing strain for each type of acoustic mode; for $\varepsilon > 0.07$ the lifetime enhancement becomes slight for ZA modes and the lifetimes of LA and TA modes decrease. These lifetime variations together with the gradual decrease of heat capacity (see Figure 2) of each type of acoustic mode reasonably explain the peak of $k$ at $\varepsilon = 0.07$. Our data analysis shows the group velocity variation plays only a minor role in determining the peak enhancement. We note that increasing strain results in decreasing anharmonic IFCs[13] which tends to increase the phonon lifetimes. At high strain the internal stress responds strongly nonlinearly to strain[35], and the decreasing magnitude of the anharmonic IFCs becomes small, resulting in only a slight enhancement of the lifetimes of ZA modes for $\varepsilon > 0.07$. For TA and LA modes, the slight reduction of anharmonic IFCs does not compensate the accumulated phonon softening at high strain, thus resulting in the decrease of lifetimes as seen for $\varepsilon = 0.1$ in Figs. 6c and 6d. For finite strained silicene with $L = 84$ μm the calculated $k$ is as high as that of bulk crystalline silicon. With increasing length, the $k$ is expected to be larger than that of silicon due to the strong size effect, showing that strained silicene may be of interest for thermal management applications in silicon-based nanoelectonics. Our calculations estimate the fracture strain of silicene is up to $\varepsilon \sim 0.16$, thus engineering $k$ by strain is promising. For stanene $k$ with $L < 20$ μm is lower than 2 W/m·K even at large strain, showing that strained stanene may be more suitable for thermoelectric applications due to its low $k$ and enhanced electronic band gap,[6,36] which benefit thermoelectric performance.

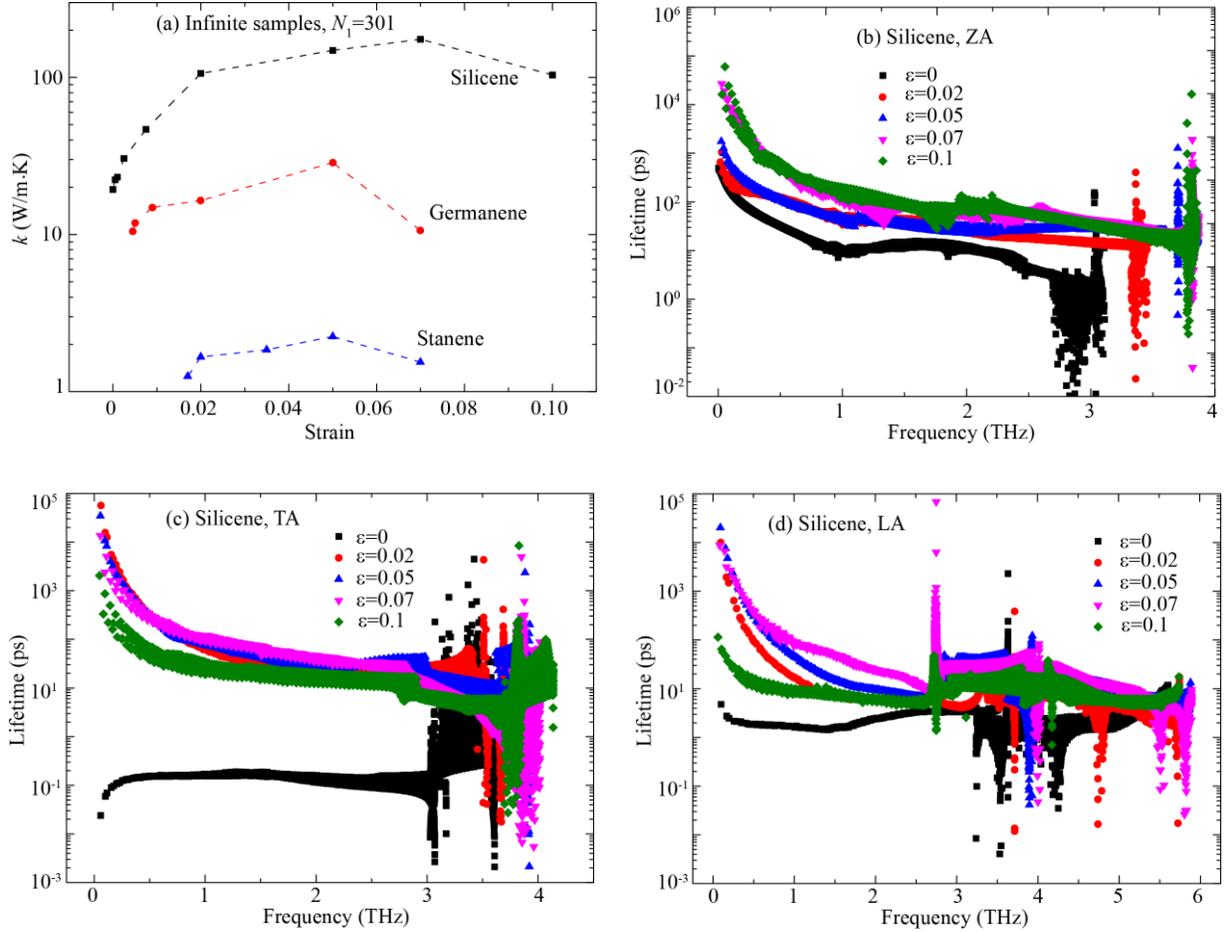

Figure 6. (a) Strain dependent room temperature $k$ for $N_1$=301 for infinite silicene, germanene and stanene. (b), (c) and (d) give room temperature lifetimes versus frequency at different strains for ZA, TA and LA modes, respectively.

In summary, we have presented a comprehensive picture of phonon thermal transport in unstrained and strained silicene, germanene and stanene based on first principles lattice dynamics calculations. Appropriate tensile strain may stabilize the periodic lattices of germanene and stanene, benefiting the fabrication and transfer of large suspended samples for applications.

Tensile strain gives a strong size effect of $k$ for these three monolayers even at high temperature. Unstrained silicene and critically strained germanene and stanene do not demonstrate this pronounced size effect. The origin of this size effect comes from nonzero contributions of ZA modes in the long wavelength limit. These ZA modes give the largest contribution to $k$ of unstrained monolayer systems compared with other modes throughout the temperature range considered. The SMRTA fails to describe phonon thermal transport in these systems. The interplay of size and strain effects leads to a peak enhancement of $k$ for each monolayer. Strained silicene may be of interest for thermal management applications due to its large $k$, up to that of bulk silicon, while strained stanene may be more suitable for thermoelectric applications due to its significantly lower $k$. These findings develop fundamental understanding of thermal transport in monolayer systems, and may guide their development for future applications.

ASSOCIATED CONTENT

**Supporting Information**

This file presents i) The calculated phonon dispersion relations of unstrained germanene with $d$-electrons in valence. ii) The self-consistent iterative process of the thermal conductivity for silicene at $\varepsilon$=0.05, germanene at $\varepsilon$=0.02 and stanene at $\varepsilon$=0.035 corresponding to $q$-point grid density $N_1$=301. This material is available free of charge via the Internet at http://pubs.acs.org.

AUTHOR INFORMATION

**Corresponding Author**

* E-mail: kuangzhang88@gmail.com.

ACKNOWLEDGMENT

We are thankful for the financial support from the Hong Kong General Research Fund under Grant No. 152140/14E and the Hong Kong Polytechnic University under Grant No. 1-99QP. L. L. acknowledges support from the U. S. Department of Energy, Office of Science, Office of Basic Energy Sciences, Materials Sciences and Engineering Division and the National Energy Research Scientific Computing Center (NERSC), a DOE Office of Science User Facility supported by the Office of Science of the U. S. Department of Energy under Contract No. DE-AC02-05CH11231.

**Supplementary Information for**

**Tensile Strains Give Rise to Strong Size Effects for Thermal Conductivities of**

**Silicene, Germanene and Stanene**

Youdi Kuang,*[†] Lucas Lindsay,[‡] Sanqiang Shi,[†] Guangping Zheng[†]

[†] Department of Mechanical Engineering, the Hong Kong Polytechnic University, Hung Hom, Kowloon, Hong Kong

[‡] Materials Science and Technology Division, Oak Ridge National Laboratory, Oak Ridge, Tennessee 37831, USA

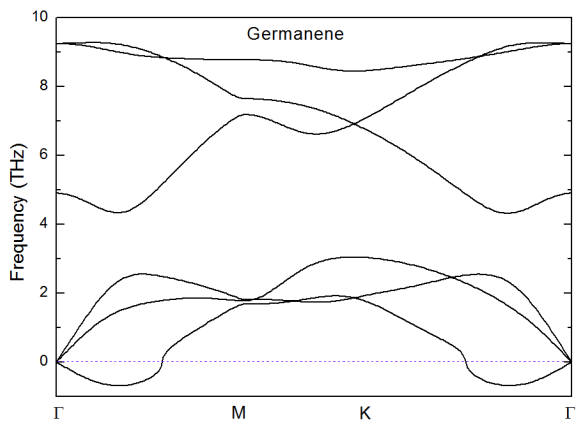

*Corresponding author. Tel.: Fax: +852-28315139. E-mail address: kuangzhang88@gmail.com(Youdi Kuang)

Figure S1. Calculated phonon dispersion of unstrained germanene (*d*-electron included in valence). The ground state was determined by full geometrical optimization based on the following setup: plane wave cutoff 560 eV, 28×28 k-mesh, energy convergence tolerance of $1.0\times10^{-9}$, force convergence precision $1.0\times10^{-6}$ eV/Å and a vacuum layer thickness 1.6 nm using the VASP package. We obtain an equilibrium lattice constant of 3.954 Å and a buckling height of 0.638 Å. The harmonic IFCs are calculated using a 7×7 supercell without additional neighbor cutoff, a 4×4 k-mesh and the same plane wave cutoff, energy convergence tolerance and vacuum layer thickness as those used for the optimization. Note the soft (imaginary) out-of-plane acoustic modes at low frequency.

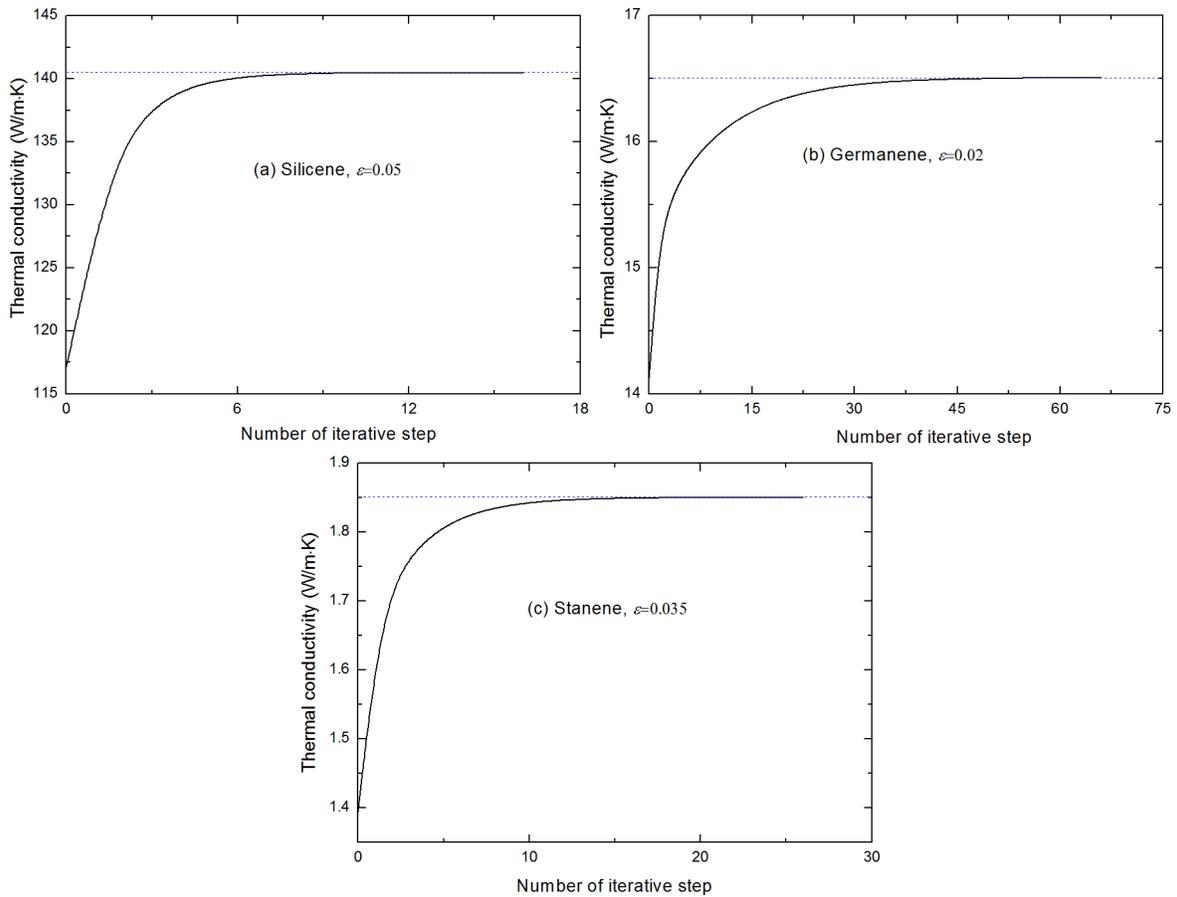

Figure S2. The thermal conductivity as a function of iterative step for silicene at $\varepsilon$=0.05, germanene at $\varepsilon$=0.02 and stanene at $\varepsilon$=0.035 corresponding to $q$-point grid density $N_1$=301. Good convergence is achieved after 16，64 and 27 iterations using convergence precision $1 \times 10^{-5}$ for silicene, germanene and stanene, respectively.